\documentclass{article}

\usepackage{graphicx}
\usepackage{hyperref}
\usepackage{amsmath}
\usepackage{amsfonts}

\newcommand{\C}{\mathcal{C}}
\newcommand{\X}{\mathbf{X}}

\newcommand{\R}{\mathbb{R}}
\newcommand{\Sc}{\mathcal{S}}
\newcommand{\I}{\mathcal{I}}

\newcommand{\task}{\texttt{revamp}}
\newcommand{\Task}{\texttt{Revamp}}

\begin{document}

\title{Model-based clustering of multi-tissue gene expression data} 

\author{Pau Erola\,$^1$, Johan LM Bj\"orkegren\,$^{2,3}$, and Tom Michoel\,$^1$ }

\date{\small{ \flushleft{$^1$ Division of Genetics and Genomics, The Roslin Institute, The University of Edinburgh, Midlothian EH25 9RG, Scotland, United Kingdom. 
\\ $^{2}$ Department of Genetics and Genomic Sciences, Institute of Genomics and Multiscale Biology, Icahn School of Medicine at Mount Sinai, NY 10029, New York, USA. 
\\ $^3$ Integrated Cardio Metabolic Centre (ICMC), Karolinska Institutet, 141 57, Huddinge, Sweden. \\ } } }

\maketitle

\abstract{\textbf{Motivation:} Recently, it has become feasible to generate large-scale, multi-tissue gene expression data, where expression profiles are obtained from multiple tissues or organs sampled from dozens to hundreds of individuals. When traditional clustering methods are applied to this type of data, important information is lost, because they either require all tissues to be analyzed independently, ignoring dependencies and similarities between tissues, or to merge tissues in a single, monolithic dataset, ignoring individual characteristics of tissues.\\
\textbf{Results:} We developed a Bayesian model-based multi-tissue clustering algorithm, \task, which can incorporate prior information on physiological tissue similarity, and which results in a set of clusters, each consisting of a core set of genes conserved across tissues as well as differential sets of genes specific to one or more subsets of tissues. Using data from seven vascular and metabolic tissues from over 100 individuals in the STockholm Atherosclerosis Gene Expression (STAGE) study, we demonstrate that multi-tissue clusters inferred by \task\ are more enriched for tissue-dependent protein-protein interactions compared to alternative approaches. We further demonstrate that \task\ results in easily interpretable multi-tissue gene expression associations to key coronary artery disease processes and clinical phenotypes in the STAGE individuals.\\
\textbf{Availability:} \Task\ is implemented in the Lemon-Tree software, available at \href{https://github.com/eb00/lemon-tree}{https://github.com/eb00/lemon-tree}.\\
\textbf{Contact:} \href{tom.michoel@roslin.ed.ac.uk}{tom.michoel@roslin.ed.ac.uk}
}

\section{Introduction}

Clustering gene expression data into groups of genes sharing the same expression profile across multiple conditions remains one of the most important methods for reducing the dimensionality and complexity of large-scale microarray and RNA-sequencing datasets \cite{dhaeseleer2005does,andreopoulos2009roadmap,vandam2017gene}. Coexpression clusters group functionally related genes together, and reveal how diverse biological processes and pathways respond to the underlying perturbation of the biological system of interest. Traditionally, clustering is performed by collecting data from multiple experimental treatments \cite{eisen1998cluster}, time points \cite{spellman1998}, cell or tissue types \cite{freeman2007construction}, or genetically diverse individuals \cite{ghazalpour2006integrating} in a single data matrix from which meaningful patterns are extracted using any of a whole range of statistical and algorithmic approaches. More recently, it has become feasible to probe systems along two or more of these dimensions simultaneously. In particular, we are interested in multi-tissue data, where gene expression profiles are obtained from multiple tissues or organs sampled from dozens to hundreds of individuals \cite{keller2008gene,hagg2009multi,greenawalt2011survey,fu2012unraveling,grundberg2012mapping,foroughi2015,franzen2016,gtex2017genetic}. These data can potentially reveal the similarity and differences in (co)expression between tissues as well as the tissue-specific variation in (co)expression across individuals. 

However, when traditional clustering methods are applied to this type of data, important information is lost. For instance, if each tissue-specific sub-dataset is clustered independently, the resulting sets of clusters will rarely align, and to compare clusters across tissues, one will be faced with the general problem of determining cluster preservation statistics \cite{langfelder2011my}. If instead the data are concatenated `horizontally' in a single gene-by-sample matrix, a common set of clusters will be found, but these will be biased heavily towards house-keeping processes that are coexpressed in all tissues. A potentially more promising approach is to concatenate data `vertically' in a tissue-gene-by-individual matrix, where the entities being clustered are `tissue-genes', the tissue-specific expression profiles of genes \cite{dobrin2009multi,talukdar2016}. However, in studies with a large number of tissues, the number of individuals with available data in \emph{all} tissues is typically very small, i.e. a large number of samples will have to be discarded to obtain a tissue-gene-by-individual matrix without missing data. 

Dedicated clustering algorithms for multi-tissue expression data are scarce, and mostly based on using the higher-order generalized singular value decomposition or related matrix decomposition techniques, to identify common and differential clusters across multiple conditions \cite{ponnapalli2011higher,li2011,xiao2014multi}. While algebraic methods are attractive for their relative efficiency and simplicity, they lack the flexibility, robustness and statistical soundness of Bayesian model-based clustering methods, which model the data as a whole using mixtures of probability distributions \cite{fraley2002model,si2013model,ickstadt2017toward}. Model-based clustering is particularly attractive for multi-tissue data, because it would allow, at least in principle, to account for different noise levels and sample sizes in different tissues and to incorporate prior information on the relative similarity between certain (groups of) tissues based on their known physiological function. Here we present a novel statistical framework and inference algorithm for model-based clustering of multi-tissue gene expression data, which can incorporate prior information on tissue similarity, and which results in a set of clusters, each consisting of a core set of genes conserved across tissues as well as differential sets of genes specific to one or more subsets of tissues.

\section{Approach}

In model-based clustering, a partitioning of genes into non-overlapping clusters parametrizes a probabilistic model from which the expression data is assumed to have been generated, typically in the form of a mixture distribution where each cluster corresponds to one mixture component. Using Bayes' theorem, this can be recast as a probability distribution on the set of all possible clusterings parameterized by the expression data, from which maximum-likelihood solutions can be obtained using expectation-maximization or Gibbs sampling \cite{fraley2002model,liu2004,ickstadt2017toward}.

Our approach to clustering multi-tissue data combines ideas from existing ordinary (``single-tissue'') \cite{qin2006,joshi2008}  and multi-species \cite{roy2013arboretum} model-based clustering  methods. We use the generative model of \cite{qin2006} and \cite{joshi2008} to obtain the posterior probability for a  (single-tissue) clustering given a (single-tisse) dataset. From \cite{roy2013arboretum} we use the idea that a multi-tissue clustering consists of a set of \emph{linked} clusters, where cluster $k$ in one tissue corresponds to cluster $k$ in any other tissue, and each cluster $k$ contains a \emph{core} set of genes, belonging to cluster $k$ in \emph{all} tissues, and a \emph{differential} set of tissue-specific genes, belonging to cluster $k$ in one or more, but not all, tissues. Like \cite{roy2013arboretum}, we assume that the data from one tissue can influence the clustering in another tissue, albeit via a simpler mechanism as we do not aim to reconstruct any phylogenetic histories among tissues. In brief, we assume that the posterior probability distribution of clusterings in tissue $t$ is given by its ordinary single-tissue distribution given the expression data for tissue $t$, multiplied by a tempered distribution for observing that same clustering given the expression data for all other tissues $t'\neq t$. The degree of tempering determines the degree of influence of one tissue on another, and can be used to model known prior relationships between tissues. For instance, we expect \textit{a priori} that coexpression clusters will be more similar between vascular tissues, than between vascular and metabolic tissues.

\section{Methods}

\subsection{Statistical model for single-tissue clustering}
\label{sec:stat-model-single}

Our method is based on previous model-based clustering algorithms applicable to single-tissue data \cite{qin2006,joshi2008}. In brief, for an expression data matrix $\X\in\R^{G\times N}$ for $G$ genes and $N$ samples, a clustering $\C$ is defined as a partition of the genes into $K$ non-overlapping sets $C_k$. We assume that the data points for the genes in each cluster and each sample are normally distributed around an unknown mean and unknown variance/precision. Given a clustering $\C$ and a set of means and precisions $(\mu_{kn},\tau_{kn})$ for each cluster $k$ and sample $n$, we obtain a distribution on expression data matrices as
\begin{equation*}
  p\bigl( \X \mid \C, \{\mu_{kn},\tau_{kn}\}\bigr) = \prod_{k=1}^K \prod_{n=1}^N \prod_{g\in C_k} p(x_{gn} \mid \mu_{kn},\tau_{kn} ).
\end{equation*}
Assuming a uniform prior on the clusterings $\C$ and independent normal-gamma priors on the normal distribution parameters, we can use Bayes' rule to find the marginal posterior probability of observing a clustering $\C$ given data $\X$, upto a normalization constant:
\begin{equation}\label{eq:1}
  P(\C\mid \X) \propto \prod_{k=1}^K \prod_{n=1}^N \iint p(\mu,\tau) \prod_{g\in C_k}  p (x_{gn} \mid \mu,\tau)\; d\mu d\tau.
\end{equation}
Note that we use a capital `$P$' to indicate that this is a \emph{discrete} distribution. $p(\mu,\tau)=p(\mu\mid\tau)p(\tau)$ is the normal-gamma prior, with
\begin{align*}
  p(\mu\mid\tau)=\bigl(\frac{\lambda_0\tau}{2\pi}\bigr)^{1/2}
  e^{-\frac{\lambda_0\tau}2 (\mu-\mu_0)^2},\quad
  p(\tau) = \frac{\beta_0^{\alpha_0}}{\Gamma(\alpha_0)}
  \tau^{\alpha_0-1} e^{-\beta_0\tau},
\end{align*}
$\alpha_0,\beta_0,\lambda_0 > 0$ and $-\infty<\mu_0<\infty$ being the parameters of the normal-gamma prior distribution.  We use the values $\alpha_0=\beta_0=\lambda_0= 0.1$ and $\mu_0=0.0$, resulting in a non-informative prior. The double integral in \eqref{eq:1} can be solved exactly in terms of the sufficient statistics $T^{(\alpha)}_{kl} = \sum_{i \in \C_k} \sum_{n=1}^N x_{in}^\alpha$ ($\alpha=0,1,2$) for each cluster, see \cite{joshi2008} for details.

For computational purposes, the decomposition of eq.~(\ref{eq:1}) into a product of independent factors, one for each cluster and sample, is important. We write the log-likelihood or Bayesian score accordingly as: 
\begin{equation}\label{eq:3}
  \Sc(\C) =\log P(\C\mid\X) = \sum_{k=1}^K \sum_{n=1}^{N} \Sc_{kn}.
\end{equation}

\subsection{Statistical model for multi-tissue clustering}
\label{sec:stat-model-multi}

Next, we assume that expression data $\X=[\X_1\in\R^{G\times N_1},\dots,\X_T\in\R^{G\times N_T}]$ is available for $G$ genes in $T$ tissues, with $N_t$ samples in each tissue $t\in\{1,\dots,T\}$. We define a multi-tissue clustering as a collection of single-tissue clusterings $\C=\{\C_1,\dots,\C_T\}$, and assume that the probability of observing $\C$ given data $\X$ is given by
\begin{multline}\label{eq:2}
  P(\C\mid \X) = P\bigl( \C_1,\dots,\C_T \mid \X_1,\dots,\X_T\bigr)\\
  = \frac1Z \prod_{t=1}^T \biggl\{ P\bigl( \C_t\mid \X_t\bigr)\, \prod_{t'\neq t} P\bigl( \C_t\mid \X_{t'}\bigr)^{\lambda_{t,t'}} \biggr\},
\end{multline}
where $Z$ is a normalization constant which we henceforth will ignore, each factor $P\bigl( \C_t\mid \X_t'\bigr)$ is a single-tissue posterior probability distribution defined in eq.~(\ref{eq:1}), and $\lambda_{t,t'}\in [0,1]$ is a set of hyper-parameters; for notational convenience we define $\lambda_{t,t}=1$.

Note that $P\bigl( \C_t\mid \X_{t'}\bigr)$ is a discrete distribution measuring how well clustering $\C_t$ is supported by data $\X_{t'}$. Raising a discrete distribution to a power less than 1 has the effect of making the distribution more uniform. Hence in eq.~(\ref{eq:2}), we are asking that clustering $\C_t$ is supported predominantly by data $\X_t$ from its own tissue, but also, albeit to a lesser extent depending on the values of $\lambda_{t,t'}$, by data from the other tissues.

Optimizing eq.~(\ref{eq:2}) across all multi-tissue clusterings is challenging. A considerable simplification is obtained if we constrain the problem to multi-tissue clusterings with the \emph{same} number of clusters $K$ in each tissue. Denoting by $\I_t$ the set of samples/individuals in tissue $t$ and by $N=\sum_{t=1}^T N_t$ the total number of samples, the decomposition in eq.~(\ref{eq:3}) allows to write:
\begin{align}
  \log  P(\C\mid \X) &= \sum_{t=1}^T \sum_{t'=1}^T \lambda_{t,t'} \log P(\C_t\mid \X_{t'}) \nonumber\\
  &= \sum_{t=1}^T \sum_{t'=1}^T \lambda_{t,t'} \sum_{k=1}^K \sum_{n\in\I_{t'}} \Sc^{(t)}_{kn}\nonumber\\
  &= \sum_{t=1}^T \sum_{k=1}^K \sum_{n=1}^{N} \gamma^{(t)}_{n} \Sc^{(t)}_{kn}, \label{eq:4}
\end{align}
where we used $\lambda_{t,t}=1$, defined $\gamma^{(t)}_n \equiv \lambda_{t,t(n)}$, with $t(n)$ the tissue to which sample $n$ belongs, and wrote $\Sc^{(t)}_{kn}$ to denote the Bayesian score of clustering $\C_t$ with respect to sample $n$.

Two extremal choices for the hyper-parameters are of interest. If $\lambda_{t,t'}=1$ for all $t,t'$, then the Bayesian score
\begin{equation}
  \label{eq:5}
  \Sc^{(t)} = \sum_{k=1}^K \sum_{n=1}^{N} \gamma^{(t)}_{n} \Sc^{(t)}_{kn}
\end{equation}
is the same for each tissue $t$ and identical to eq.~(\ref{eq:3}) for the concatenated data matrix $\X=[\X_1,\dots,\X_T]$. Hence this is  equivalent to clustering the entire dataset as if it came from a single-tissue (termed `horizontal' data concatenation in the Introduction). If $\lambda_{t,t'}=0$ for $t'\neq t$, then eq.~(\ref{eq:2}) decomposes as a product of independent single-tissue factors. This is equivalent to clustering each tissue sub-dataset independently.

\subsection{Optimization algorithm}
\label{sec:optim-algor}

To find a local maximum of the Bayesian score in eq.~(\ref{eq:4}), the following heuristic, greedy optimization algorithm was used:
\begin{enumerate}
\item \textbf{Data standardization:} Using appropriately normalized gene expression data, each gene is standardized to have mean zero and standard deviation one on the concatenated data $\X$.
\item \textbf{Determine the number of clusters:} K-means clustering is run on the concatenated data with the number of clusters ranging from 2 to 100. The optimal number $K$ is selected by visual inspection of an elbow plot.
\item \textbf{Initialize multi-tissue clustering:} Starting from the k-means clustering output at the selected number of clusters, genes are reassigned until a local optimum is reached for the single-tissue score eq.~(\ref{eq:3}) on the concatenated data $\X$. All $\C_t$ are initialized by this clustering.
\item \textbf{Optimize multi-tissue clustering:} For each tissue $t$, optimize $\C_t$ by finding a local maximum for the Bayesian score eq.~(\ref{eq:5}) using single-gene reassignments; only gene reassignments improving the score by a minimum threshold $\epsilon$ are considered.
\end{enumerate}
Note that even in the case $\lambda_{t,t'}=0$ for $t'\neq t$, which removes all tissue dependencies in the Bayesian score (\ref{eq:4}), this algorithm still results in a multi-tissue clustering with linked clusters, due to each tissue being initialized by the same clustering and converging to a local optimum.

\subsection{Implementation}
\label{sec:implementation}

The statistical model and optimization algorithm have been implemented in java, as an extension of the `task' \task\ in the Lemon-Tree software, available at \href{https://github.com/eb00/lemon-tree}{https://github.com/eb00/lemon-tree}.

\subsection{The STockholm Atherosclerosis Gene Expression study}
\label{sec:stage}

In the STockholm Atherosclerosis Gene Expression (STAGE) study, 612 tissue samples from 121 individuals were obtained during coronary artery bypass grafting surgery from the atherosclerotic arterial wall (AAW, $n=73$), internal mammary artery (IMA, $n = 88$), liver ($n = 87$), skeletal muscle (SM, $n = 89$), subcutaneous fat (SF, $n = 72$) and visceral fat (VF, $n = 98$) of well-characterized CAD patients; fasting whole blood (WB) was obtained for isolation of DNA ($n = 109$) and RNA ($n = 105$) and biochemical analyses. RNA samples were used for gene expression profiling with a custom Affymetrix array (HuRSTA- 2a520709) \cite{hagg2009multi,foroughi2015,talukdar2016}. To enable comparison across tissues, all samples were jointly normalized using Affymetrix Power Tools (v1.4.2). 4956 genes with variance greater than 1 across all 612 samples were selected for further analysis, and subsequently standardized to have mean zero and standard deviation one, again across all 612 samples. 

\subsection{Methods for comparison}
\label{sec:methods-comparison}

We ran three multi-tissue clustering methods:
\begin{itemize}
\item \Task\ with reassignment threshold $\epsilon=$0.005 and prior tissue similarities $\lambda_{t,t'}=\rho_{t,t'}^\alpha$, where $\rho_{t,t'}$ is the average correlation coefficient between samples from tissue $t$ and $t'$ measured in the same individual and $\alpha=0.25$.
\item \Task\ with reassignment threshold $\epsilon=$0.005 and prior tissue similarities $\lambda_{t,t'}=0$.
\item An alternative method, which treats the expression profile of each gene $g$ in each tissue $t$ as a separate (gene, tissue) variable and clusters the resulting (gene,tissue)-by-individual expression matrix using the single-tissue clustering algorithm (Section \ref{sec:stat-model-single}). This results in a single set of clusters, which are disentangled into a set of linked clusters, by assigning gene $g$ to cluster $m$ in tissue $t$ whenever $(g,t)$ belongs to original cluster $m$. This method was called ``vertical data concatenation'' before, and relies on having expression data from multiple tissues in the \emph{same} individual. In STAGE, 21 individuals had data in all 7 tissues.
\item Single-tissue clustering on the entire dataset of 612 samples (called ``horizontal data concatenation'' before). This results in an identical clustering across all tissues. It is not a true multi-tissue clustering method, but is used as an overall benchmark to determine the relevance of a multi-tissue approach.
\end{itemize}

\subsection{Validation data}
\label{sec:validation-data}

To evaluate the biological relevance of each multi-tissue clustering method, we used the following approach:
\begin{itemize}
\item We performed GSEA using first the GOSlim ontology, that gives a broad overview of the ontology content without the detail of the specific fine-grained terms (\href{http://www.geneontology.org/page/go-slim-and-subset-guide}{http://www.geneontology.org/page/go-slim-and-subset-guide}), and after on GO terms (\href{http://www.geneontology.org/page/download-ontology}{http://www.geneontology.org/ page/download-ontology}).
\item We assigned sets of ``regulators'' to each of the modules considering as candidate regulators the tissue-specific sets of genes with significant eQTLs (2464 AAW, 3209 IMA, 4491 liver, 2534 SM, 2373 SF, 2994 VF and 5691 WB genes) \cite{foroughi2015}.
\item We obtained human tissue protein--protein interaction (PPI) networks from \cite{barshir2012tissuenet}. Specifically, we used TissueNet v2 networks consisting of curated experimentally detected PPIs between proteins expressed in Genotype-Tissue Expression dataset tissues `Artery -- Aorta', `Liver', `Muscle -- Skeletal', `Adipose -- Subcutaneous', `Adipose -- Visceral (Omentum)' and `Whole Blood', available for download at \href{http://netbio.bgu.ac.il/labwebsite/?q=tissuenet2-download}{http://netbio.bgu.ac.il/labwebsite/?q=tissuenet2-download}.
\end{itemize}

\subsection{Validation methods}
\label{sec:validation-methods}

We tested for GO functional enrichment using the task \texttt{go\_annotation} in the Lemon-Tree software,
and task  \texttt{regulators}was used to identify gene ``regulators'' using a probabilistic scoring, taking into account the profile of the candidate regulator and how well it matches the profiles of co-expressed genes \cite{joshi2009}.

To test for enrichment of known PPIs in a given clustering, we calculated the fold-change enrichment as
\begin{align*}
  \text{FC} = \frac{\frac{\text{Number of co-clustered gene pairs with PPI}}{\text{Total number of PPI}}} 
  {\frac{\text{Total number of co-clustered gene pairs}}{\text{Total number of gene pairs}}}
\end{align*}
All clustering methods were run on the seven available STAGE tissues, and the results for six tissues were used for validation (IMA did not have a matching tissue in the TissueNet database). To evaluate the clustering of a particular tissue, we used all PPIs for that tissue. To evaluate the core gene set of a cluster (for cluster $m$, the set of genes belonging to $m$ in all tissues), we used the set of PPIs shared across all tissues.

Because the fold-change value is influenced by the number of clusters (more clusters results in fewer co-clustered pairs), we used the same number ($k=12$) of clusters for all compared methods (Section~\ref{sec:methods-comparison}).

\section{Results}

\subsection{Independent clustering of multi-tissue data does not capture biological information shared across tissues}
\label{sec:indep-clust-multi}

\begin{figure}
  \centering
  \includegraphics[width=0.9\linewidth]{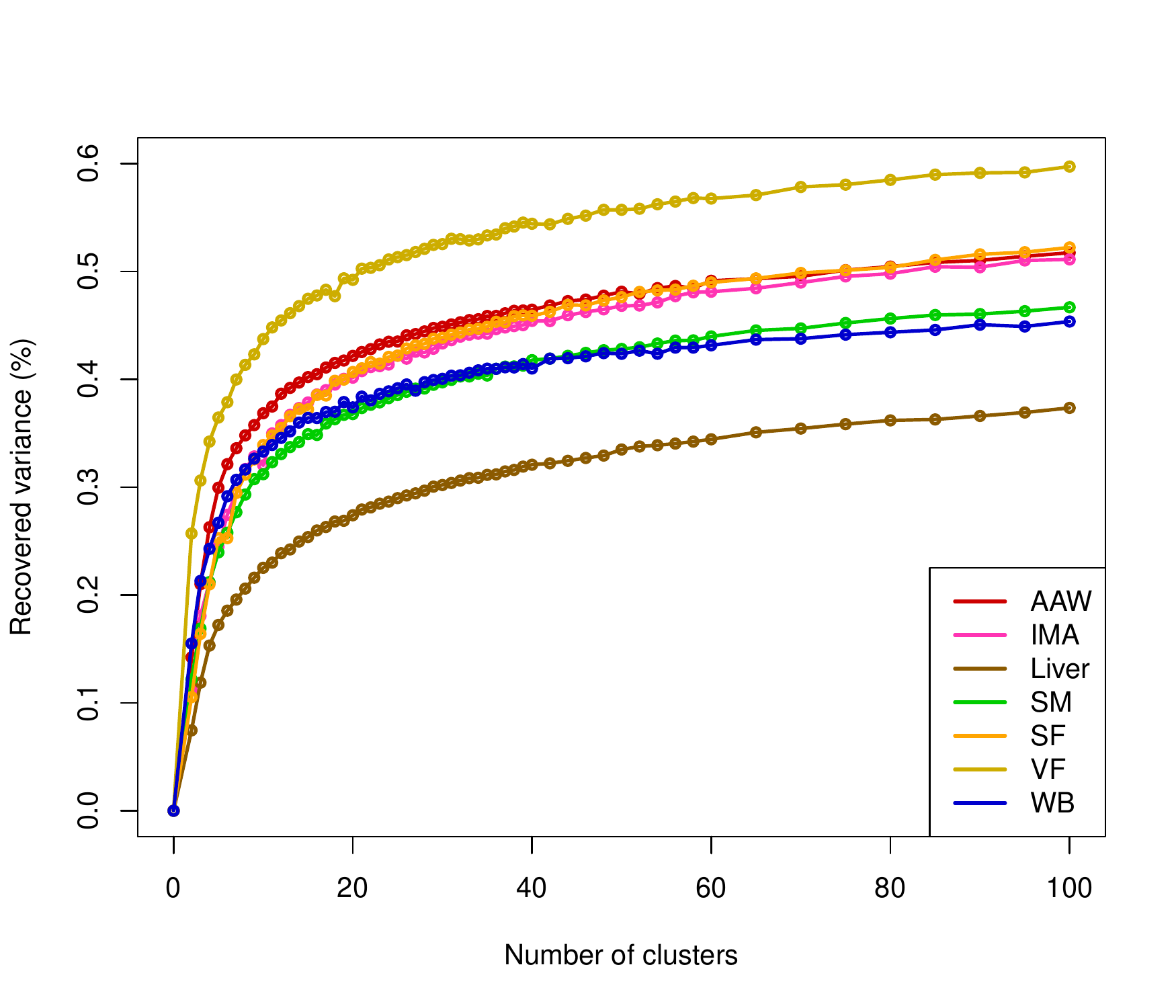}
  \caption{Percentage of variance explained  as a function of the number of clusters ($k=2,3,...100$) in the partitions obtained with k-means algorithm.}
  \label{fig:elbow}
\end{figure}

\begin{figure}
  \centering
  \includegraphics[width=0.9\linewidth]{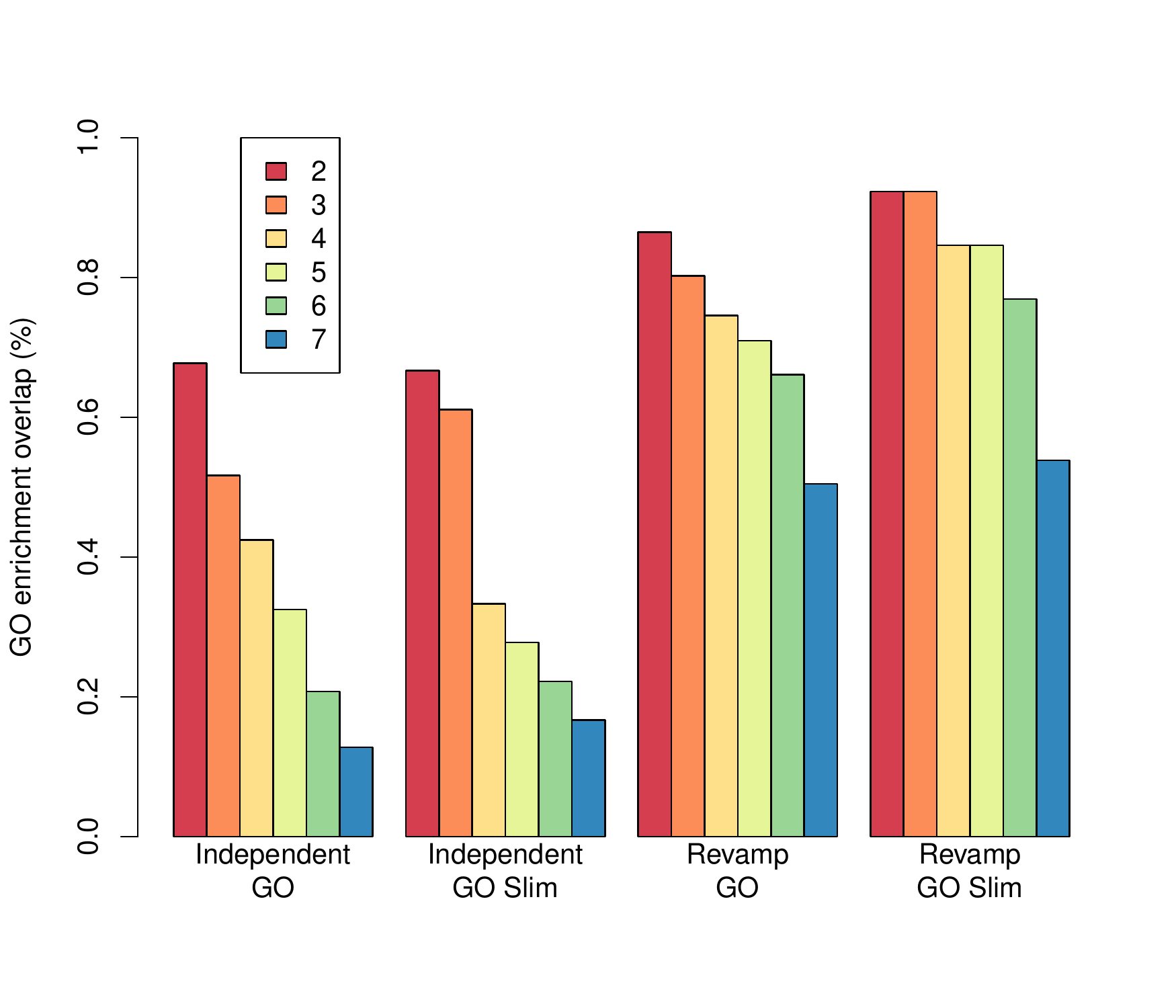}
  \caption{Percentage of shared enriched GO terms between different number of tissues for independent clustering and the proposed multi-tissue clustering method.}
  \label{fig:barplot_shared}
\end{figure}

\begin{figure}
  \centering
  \includegraphics[width=\linewidth]{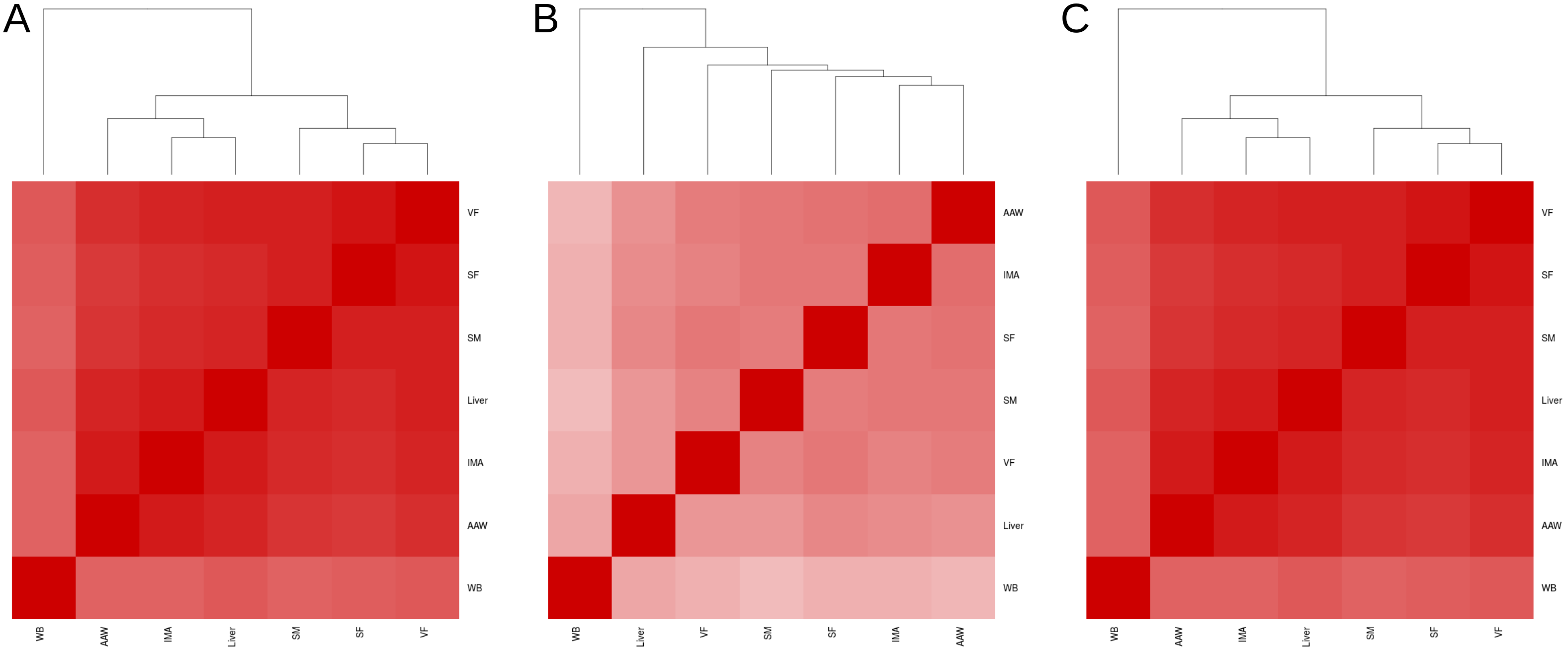}
  
  \includegraphics[width=\linewidth]{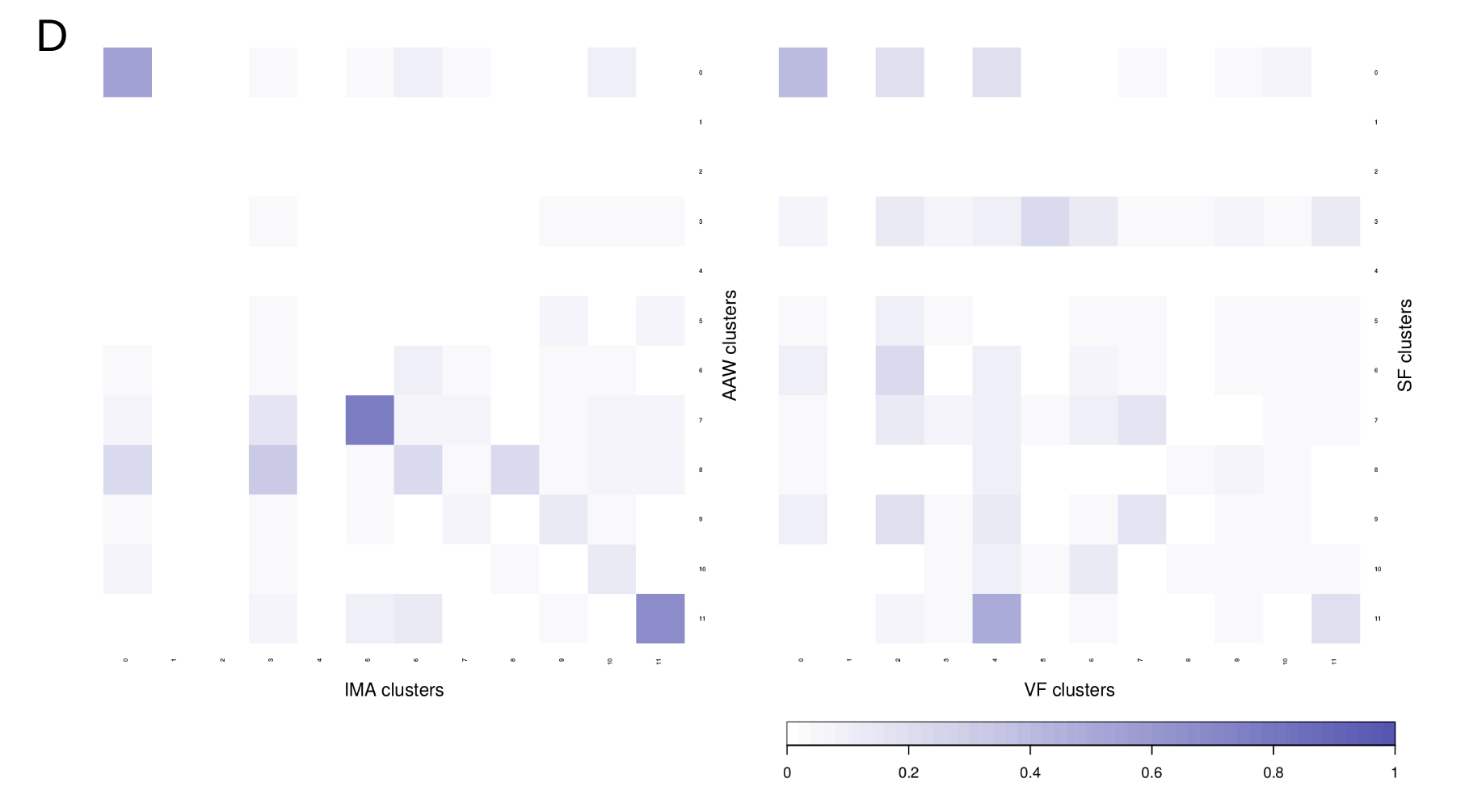}
  \caption{On the top, the correlation matrix of gene expression values between tissues (A) and the matrix of shared GO enrichments between tissues obtained with the clustering the tissues independently (B) and with our method \task\ (C). On the bottom, percentage of genes that are present in the same cluster both vascular (AAW, IMA) and adipose (SF, VF) tissues after clustering tissues independently (D).}
  \label{fig:hierarchical}
\end{figure}

To study the necessity of dedicated multi-tissue clustering algorithms, we analyzed 612 samples from 7 tissues from 121 individuals in the STockholm Atherosclerosis Gene Expression (STAGE) study \cite{hagg2009multi,foroughi2015,talukdar2016}. We clustered data from each tissue independently using k-means and selected $k=12$ clusters for all our analyses, as this value was near the inflection point of the elbow plots in all tissues as it is shown in Fig.~\ref{fig:elbow}. We then tested all 12 clusters in all tissues for functional enrichment using (fine-grained) GO and (coarse-grained) GOSlim categories. We found that an important proportion of both GO and GOSlim enriched categories were shared across pairs of tissues, but this value falls below 40\% when we compare more than 3 tissues (Fig.~\ref{fig:barplot_shared}). Moreover, similarity heatmaps (Fig.~\ref{fig:hierarchical}A-B) showed that the degree of shared enrichment between tissues didn't reflect the degree of overall expression similarity.
The similarities between tissues are also poorly reflected at the level of individual clusters. In particular, the cluster assigments between even the most closely related tissue pairs (AAW/IMA and SF/VF) did not show any clear pattern to allow to map clusters identified in one tissue to clusters identified in other tissues Fig.~\ref{fig:hierarchical}D.

\subsection{Multi-tissue clustering with \task\ produces mappable clusters with tunable overlap levels} 
\label{sec:multi-tiss-clust}

To identify co-expression clusters that reflect biological similarities and differences across tissues more accurately, we developed a model-based multi-tissue clustering algorithm \task, and implemented it in the Lemon-Tree software \cite{bonnet2015}. \Task\ identifies a set of linked clusters, where cluster $m$ in one tissue corresponds to cluster $m$ in any other tissue. Each linked cluster is anchored on a core set of genes that are co-expressed and shared across all tissues, with additional sets of genes that co-express with the core set in a tissue-specific manner (see Approach and Methods). Unlike other algorithms \cite{xiao2014multi}, \task\ can model non-linear relationships between genes and incorporate prior information on the expected similarities between tissues. \Task\ starts from an initial k-means clustering based on all tissue samples (or a representative randomly selected subset), and then updates the cluster assignments for each tissue independently using a Bayesian model-based score that depends on two (sets of) hyper-parameters. A \emph{reassignment threshold} parameter is used to control the reassignment of genes from one cluster to another in the greedy optimization algorithm, by only accepting such reassignments if they increase the Bayesian score by more than the threshold value. A set of \emph{tissue similarity} parameters with values between zero and one is used to control the degree of influence of expression data from one tissue on the cluster reassignment in another tissue.

\begin{figure}
  \centering
  \includegraphics[width=\linewidth]{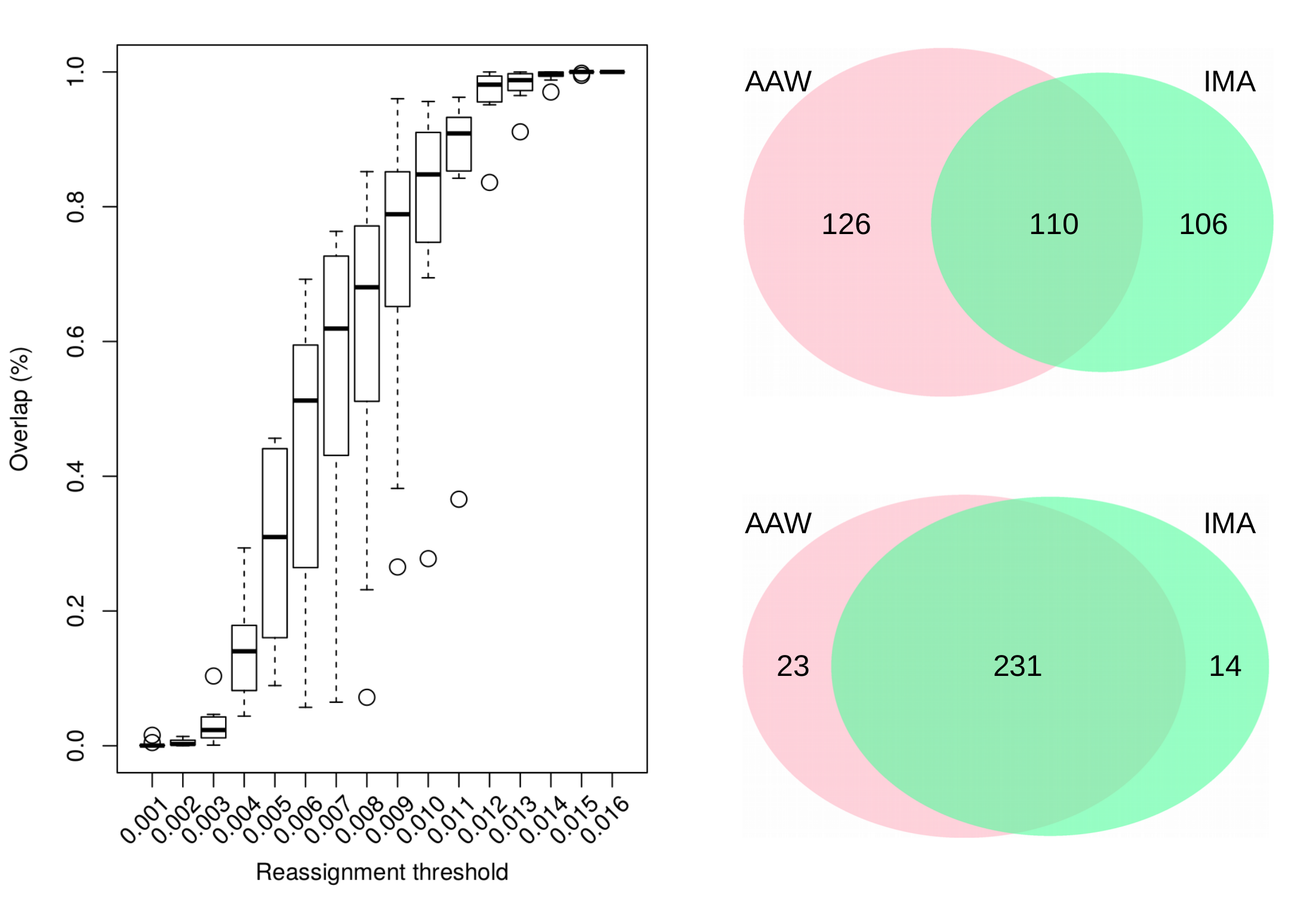}
  \caption{On the left, average percentage of genes that overlap in each cluster as function of the reassignment parameter. On the right, Venn diagram that compares the set of genes in one cluster for AAW and IMA using two threshold parameters $\epsilon=$0.001  (top) and $\epsilon=$0.005 (bottom).}
  \label{fig:boxplot_thr}
\end{figure}

To test the influence of these parameters on the clustering outcome, we again used $k=12$ for the initial k-means clustering and systematically tested (Fig.~\ref{fig:boxplot_thr}) a large space of parameter combinations for the subsequent optimization algorithm. Both the reassignment threshold and tissue similarities ultimately govern the degree of overlap across tissues of the linked clusters, with small thresholds and near-zero similarities leading to nearly tissue-independent clusterings, and large thresholds and/or near-one similarities leading to nearly identical clusterings. Although the reassignment and prior tissue similarity parameters are to some extent interchangeable (i.e.\ a smaller threshold value can be compensated by a uniform increase in similarity values),  setting the reassignment threshold to a small, non-zero value is recommended to avoid spurious reassignments due to numerical round-off errors in the Bayesian score calculation.

\subsection{\Task\ multi-tissue clustering is more enriched for tissue protein--protein interactions than other approaches}
\label{sec:revamp-multi-tissue}

To evaluate the performance of \task, we ran four multi-tissue clustering methods: \task\ with prior tissue similarities set according to their overall gene expression similarity, \task\ with prior tissue similarities set to zero, single-tissue clustering on vertically concatenated data, i.e. on (gene,tissue)-by-individual data, and single-tissue clustering on horizontally concatenated data, i.e. on gene-by-sample data, where each sample is from a particular individual in a particular tissue, see Methods for details. For each method, we tested for enrichment of human tissue protein-protein interactions (PPIs) from the TissueNet database  \cite{barshir2012tissuenet} among co-clustered genes, using six tissues that matched between STAGE and TissueNet (see Methods).

\begin{figure}
  \centering
  \includegraphics[width=0.9\linewidth]{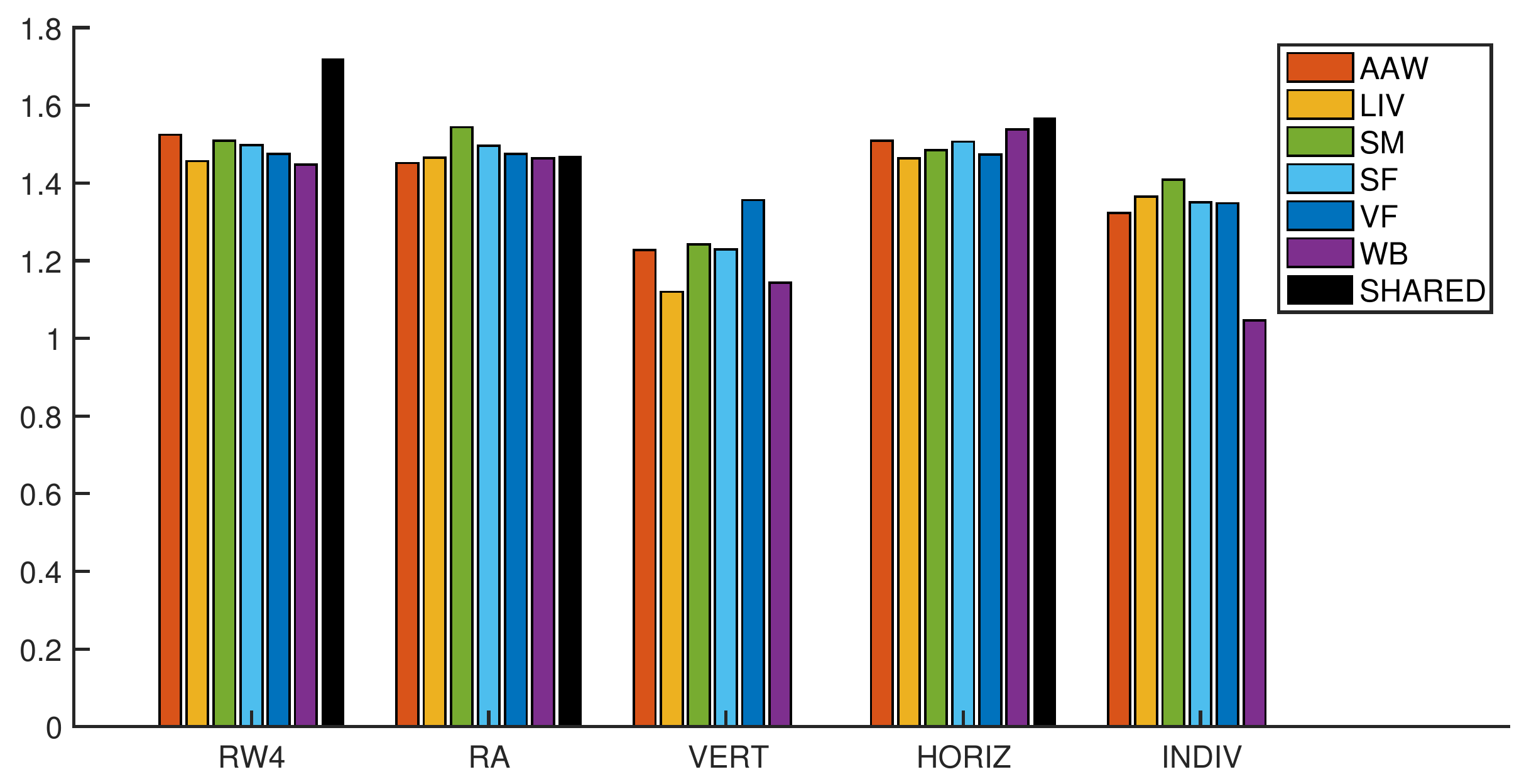}
  \caption{Fold-change enrichment of tissue PPIs in tissue clusters for four multi-tissue clustering methods. RW4 -- \task\ with prior tissue similarities set according to their overall expression correlation, RA -- \task\ with prior tissue similarities set to zero, VERT -- vertical data concatenation, HORIZ -- horizontal data concatenation, INDIV -- each tissue clustered individually. Each colored bar shows the fold-change overlap of tissue PPIs in clusters for the matching tissue; the black bar shows the fold-change overlap of tissue-shared PPIs in tissue-shared genes of linked clusters. See Methods for details.}
  \label{fig:overlap-tissuenet}
\end{figure}

On a tissue-by-tissue basis, running \task\ with or without prior tissue similarity values resulted in similar fold-change enrichment values for tissue PPIs  (average fold-change over 6 tissues of 1.49 and 1.48, respectively) as running single-tissue clustering on all 612 STAGE samples together (average fold-change 1.50), and considerably higher enrichment than using vertically concatenated data (average fold-change 1.22) (Fig.~\ref{fig:overlap-tissuenet}). For a baseline reference, we also calculated enrichment for each tissue clustered individually using the single-tissue clustering method. Consistent with the assumption that analyzing data integratively using multi-tissue clustering should improve biological relevance, single-tissue clustering resulted in lower fold-change values (average fold-change 1.31) (Fig.~\ref{fig:overlap-tissuenet}).

We further reasoned that genes assigned consistently to the same cluster across all tissues (`core' cluster genes) should reflect tissue-independent interactions between these genes. To test this hypothesis, we calculated enrichment of tissue-independent PPIs (i.e.\ PPIs present in all six tissue PPI networks) among core cluster genes. For \task\ with prior tissue similarity values, a significant increase in enrichment for  tissue-independent PPIs was observed (fold-change 1.72), whereas for \task\ without prior tissue similarities and horizontal data concatenation no difference was observed compared to all tissue PPIs (fold-changes 1.47 and 1.57, respectively) (Fig.~\ref{fig:overlap-tissuenet}). Vertical data concatenation resulted in very small core gene sets, containing no known tissue-independent PPIs.

\subsection{Functional predictions by \Task\ clusters and gene regulators associated with CAD}
\label{sec:revamp-disc-multi}

To test whether the clustering algorithm accurately captures the higher-level biological process represented by each module, or set of processes, we first performed gene ontology enrichment (Lemon-Tree task \texttt{go\_annotation}) using the GO Slim ontology. Most significant enrichments are summarized in Table~\ref{tab:goslim}. 
Network analysis revealed three connected components: clusters 5, 9 and 10 were related with immune system response; the lipid metabolic process was enriched in clusters 4, 6 and 7; and clusters 0 and 8 were associated with cell adhesion and extracellular matrix organization.

\begin{table}
  \centering
\begin{tabular}{l r r}
\hline
\textbf{GO term} & \textbf{cluster} &  \multicolumn{1}{c}{\textbf{p-val range}}  \\
\hline
immune system process & 5,9,10 & 1.04E-19 -- 3.00E-5\\
signal transduction & 9 & 1.28E-2 -- 3.82E-2 \\
macromolecular complex assembly & 10 & 1.13E-2 -- 4.88E-2\\
carbohydrate metabolic process & 10 & 1.40E-2 -- 3.82E-2\\
protein complex assembly & 10 & 1.22E-2 -- 4.88E-2\\
cellular component assembly & 10 & 1.22E-2 -- 4.88E-2\\
\hline
lipid metabolic process & 4,6,7 & 6.26E-4 -- 1.89E-2\\
transmembrane transport & 6 & 2.98E-3 -- 2.98E-3 \\
\hline
cell adhesion & 0,8 & 2.54E-9 -- 9.57E-5\\
extracellular matrix organization & 0,8 & 6.24E-5 -- 1.84E-2\\
cell proliferation & 0 & 2.73E-3 -- 2.54E-2\\
cell-cell signaling & 8 & 2.95E-3 -- 3.33E-2\\
\hline
& &
\end{tabular}
  \caption{GO functional enrichment using GOSlim ontology. The terms presented are associated with all seven STAGE tissues.}
  \label{tab:goslim}
\end{table}

\begin{figure*}
  \centering
  \includegraphics[width=\linewidth]{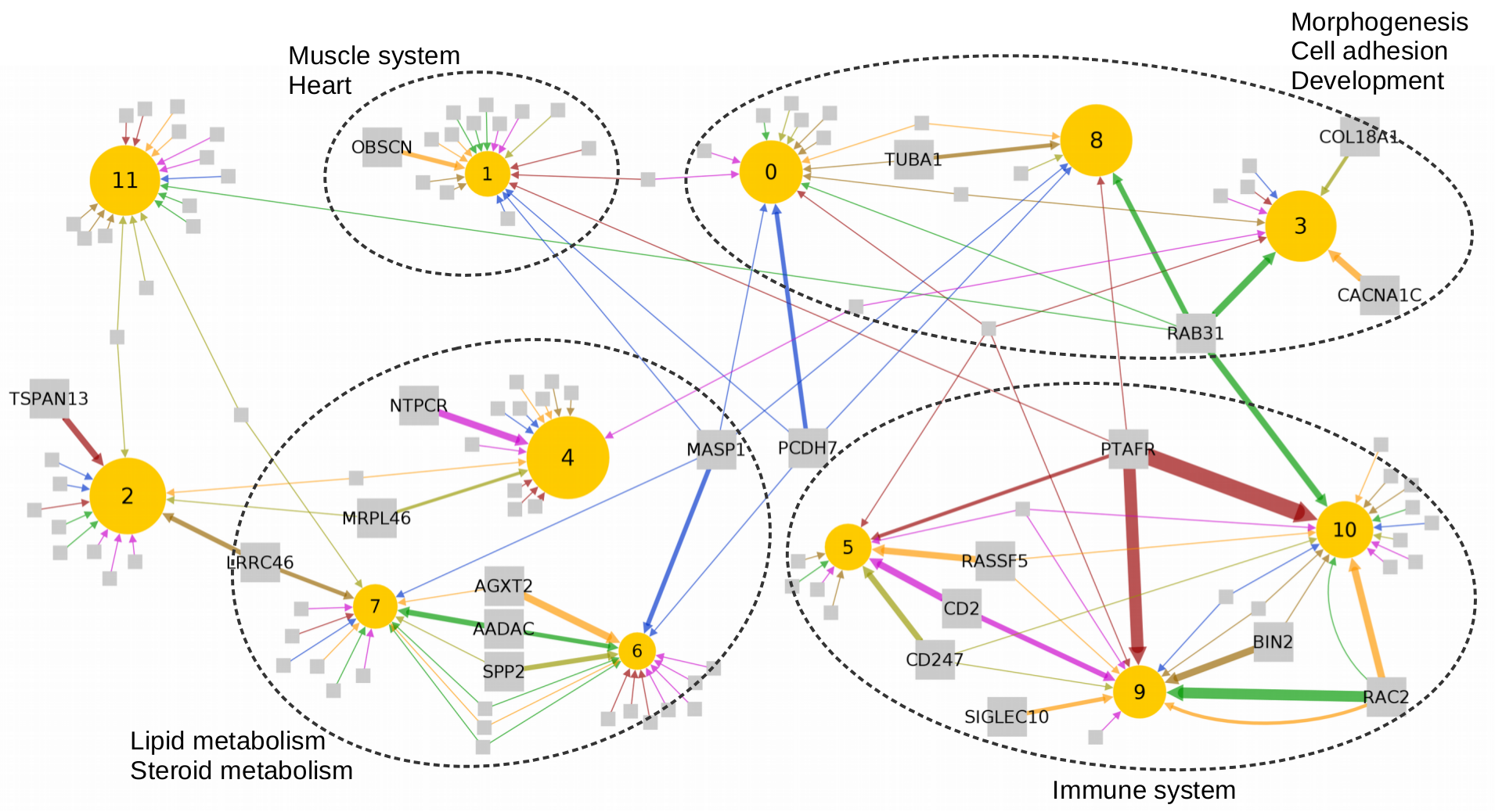}
  \caption{Module regulatory network for all 7 tissues. Only the regulators with a score greater than 20 in the \texttt{regulators} task are represented, and we named those with a score above 60. Regulators are presented as squares and clusters as circles with size proportional to the number of genes in the cluster. Edges are coloured per tissue as per Fig.~\ref{fig:net-phenotype}, and their width is proportional to the regulator score.  }
  \label{fig:net-regulators}
\end{figure*}

Then we ran independently on each tissue for each cluster the regulator probabilistic scoring task (Lemon-Tree task \texttt{regulators}) to predict upstream regulatory genes, considering as candidate regulators the tissue-specific genes with genetic variants in their regulatory regions affecting gene expression (``\textit{cis}-eQTL effects'') in the STAGE individuals \cite{foroughi2015} .
The regulatory network of the most significant regulators for the inferred modules is depicted in Fig.~\ref{fig:net-regulators}.

The development of atherosclerosis is in large part mediated by the inflammatory cascade \cite{Crowther2005}.
Our results indicated that the inflammatory response in AAW may be regulated by 
PTAFR, a mediator in platelet aggregation and the inflammatory response \cite{Perisic2016, Rastogi2008}.
SF and VF were shown to be regulated by SIGLEC10 and CD247, respectively, genes that have been previously associated with CAD
\cite{Ammirati2008, Shen2013}.
Other tissues were linked to the previously identified inflammatory regulators BIN2 \cite{Liao2011},
CD2 \cite{Hansson2006},
RAC2, that also directs plaque osteogenesis \cite{Ceneri2017},
and 
the pro-apoptotic regulator of RAS protein, RASSF5 \cite{Dejeans2010}.

Lipid metabolism also plays a key role in the development of atheroma plaques.
Metabolism-related clusters 6 and 7 were found to be regulated by genes AGXT2 and SPP2, in SF and VF respectively.
AGXT2 polymorphisms were identified as risk for CAD in Asian populations \cite{Zhou2014, Yoshino2014},
and SPP2 may contribute to the atheroprotective effects of HDL \cite{Abdel-Latif2015}.
These clusters may also be regulated by AADAC in SM, that
controls the export of sterols \cite{Tiwari2007}.
In WB, we found MASP1, a gene associated with a decreased lectin pathway activity in acute myocardial infarction patients \cite{Yan2016}.

The atherogenic pathway involves the inflammation of the arterial wall, the injury of the intima, lipid infiltration and the activation of the angiogenic signaling, processes that involve a dysfunction in the cell adhesion \cite{Sun2014}.
Our analysis showed that RAB31, which induces lipid accumulation in atheroma plaques \cite{Fu2002},
regulates the morphogenesis-related clusters 3 and 8 in SM.
Cluster 3 was also shown to be regulated by CACNA1C in SF, a gene involved in calcium channels and associated with inherited cardiac arrhythmia \cite{Kawashiri2014},
and COL18A1 in VF, that may control angiogenesis and vascular permeability \cite{Moulton1999}.
The expression levels of the potential regulators 
PCDH7, gene involved in cell adhesion, 
and TUBA1 were previously correlated with the extent of CAD \cite{Eyster2011, Sinnaeve2009, Chittur2008}.

\subsection{\Task\ discovers multi-tissue clusters underlying CAD phenotypes}
\label{sec:revamp-phenotype}

\begin{figure*}
  \centering
  \includegraphics[width=\linewidth]{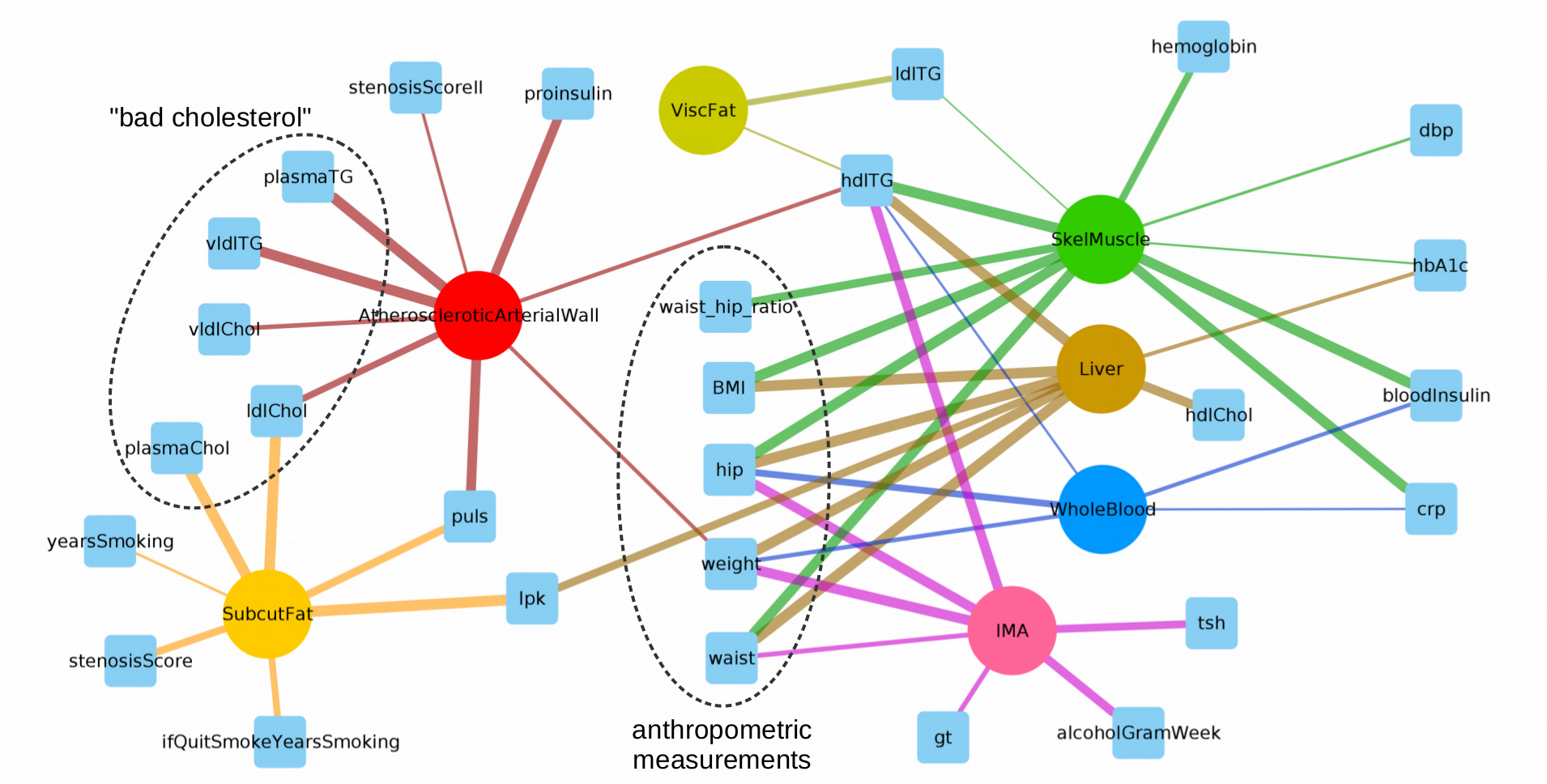}
  \caption{Network representation of the correlation between the eigengenes, the first principal component of a given module, and relevant CAD phenotypes (squares), aggregated per tissue (circles). Edge width is inversely proportional to the correlation p-value.}
  \label{fig:net-phenotype}
\end{figure*}

The systems genetics paradigm says that genetic variants in regulatory regions affect nearby gene expression (``\textit{cis}-eQTL effects''), which then causes variation in downstream gene networks (``\textit{trans}-eQTL effects'') and clinical phenotypes.
We therefore used regression analysis to identify associations between module gene expression and CAD phenotypes (see \cite{talukdar2016}), as presented in Fig.~\ref{fig:net-phenotype}.

The aggregated results revealed that AAW and SF are the main tissues associated with very-low-density lipoprotein (VLDL) and low-density lipoprotein (LDL) cholesterol levels, while the liver was the main tissue associated with high-density lipoprotein (HDL) cholesterol.
Besides that, IMA was found to be associated in cluster 3 with the thyroid-stimulating hormone, that causes many hemodynamic effects and influences the structure of the heart and circulatory system \cite{Grais2014},
and alcohol consumption in clusters 5 and 9, whose associations with cardiovascular diseases are heterogeneous \cite{Bell2017}.

On the other hand, the results showed that the phenotypes related to anthropometric measurements are mostly associated with SM, liver and IMA, and with less significance with WB and AAW but they are not correlated with the SF and VF.
If we focus on clusters related to body weight, as a typical example of a trait regulated by, and affecting multiple tissues, we can find gene regulators such as PTAFR (in AAW) and CD2 (in IMA) which have been described to affect food intake and body weight, apart from the inflammatory response \cite{Li2015, AreHanssen2004}.
In SM, RAC31 may influence on the body weight by mediating the insulin-stimulated glucose uptake \cite{Lyons1999}. Last, also the candidate regulators BIN2 and RAC2 have been associated with obesity and metabolic syndrome \cite{Aguilera2013, Zhang2005}.

\section{Conclusion}

Herein we proposed a Bayesian model-based multi-tissue clustering algorithm, \task, which incorporates prior information on physiological tissue similarity, and which results in a set of clusters consisting of a core set of genes conserved across tissues as well as differential sets of genes specific to one or more subsets of tissues. Using data from seven vascular and metabolic tissues from over 100 individuals in the STockholm Atherosclerosis Gene Expression (STAGE) study, we demonstrated that our method resulted in multi-tissue clusters with higher enrichment of tissue-specific protein-protein interactions than comparable clustering algorithms. The multi-tissue clusters inferred from the STAGE data highlighted the ability of \task\ to link together regulatory genes, biological processes and clinical patient characteristics in a meaningful way across multiple tissues, and we believe this makes it an attractive and statistically sound method for analyzing multi-tissue gene expression datatsets in general. \Task\ is implemented and freely available in the Lemon-Tree software, at \href{https://github.com/eb00/lemon-tree}{https://github.com/eb00/lemon-tree}.

% \section*{Acknowledgements}

\section*{Funding}

This work has been supported by funding from the BBSRC [Roslin Institute Strategic Programme, BB/P013732/1] and the NIH [NHLBI R01HL125863].\vspace*{-12pt}

\bibliographystyle{plain}
\bibliography{bibsysbiol}

\end{document}